\documentstyle[twoside,epsfig]{article}

\catcode`\@=11
\long\def\@makefntext#1{
\protect\noindent \hbox to 3.2pt {\hskip-.9pt  
$^{{\eightrm\@thefnmark}}$\hfil}#1\hfill}		

\def\@makefnmark{\hbox to 0pt{$^{\@thefnmark}$\hss}}	
	
\def\ps@myheadings{\let\@mkboth\@gobbletwo
\def\@oddhead{\hbox{}
\rightmark\hfil\eightrm\thepage}   
\def\@oddfoot{}\def\@evenhead{\eightrm\thepage\hfil
\leftmark\hbox{}}\def\@evenfoot{}
\def\sectionmark##1{}\def\subsectionmark##1{}}

\oddsidemargin=\evensidemargin
\newcounter{sectionc}\newcounter{subsectionc}\newcounter{subsubsectionc}
\renewcommand{\section}[1] {\vspace{12pt}\addtocounter{sectionc}{1} 
\setcounter{subsectionc}{0}\setcounter{subsubsectionc}{0}\noindent 
	{\tenbf\thesectionc. #1}\par\vspace{5pt}}
\renewcommand{\subsection}[1] {\vspace{12pt}\addtocounter{subsectionc}{1} 
	\setcounter{subsubsectionc}{0}\noindent 
	{\bf\thesectionc.\thesubsectionc. {\kern1pt \bfit #1}}\par\vspace{5pt}}
\renewcommand{\subsubsection}[1] {\vspace{12pt}\addtocounter{subsubsectionc}{1}
	\noindent{\tenrm\thesectionc.\thesubsectionc.\thesubsubsectionc.
	{\kern1pt \tenit #1}}\par\vspace{5pt}}
\newcommand{\nonumsection}[1] {\vspace{12pt}\noindent{\tenbf #1}
	\par\vspace{5pt}}

\newcounter{appendixc}
\newcounter{subappendixc}[appendixc]
\newcounter{subsubappendixc}[subappendixc]
\renewcommand{\thesubappendixc}{\Alph{appendixc}.\arabic{subappendixc}}
\renewcommand{\thesubsubappendixc}
	{\Alph{appendixc}.\arabic{subappendixc}.\arabic{subsubappendixc}}

\renewcommand{\appendix}[1] {\vspace{12pt}
        \refstepcounter{appendixc}
        \setcounter{figure}{0}
        \setcounter{table}{0}
        \setcounter{lemma}{0}
        \setcounter{theorem}{0}
        \setcounter{corollary}{0}
        \setcounter{definition}{0}
        \setcounter{equation}{0}
        \renewcommand{\thefigure}{\Alph{appendixc}.\arabic{figure}}
        \renewcommand{\thetable}{\Alph{appendixc}.\arabic{table}}
        \renewcommand{\theappendixc}{\Alph{appendixc}}
        \renewcommand{\thelemma}{\Alph{appendixc}.\arabic{lemma}}
        \renewcommand{\thetheorem}{\Alph{appendixc}.\arabic{theorem}}
        \renewcommand{\thedefinition}{\Alph{appendixc}.\arabic{definition}}
        \renewcommand{\thecorollary}{\Alph{appendixc}.\arabic{corollary}}
        \renewcommand{\theequation}{\Alph{appendixc}.\arabic{equation}}
        \noindent{\tenbf Appendix \theappendixc #1}\par\vspace{5pt}}
\newcommand{\subappendix}[1] {\vspace{12pt}
        \refstepcounter{subappendixc}
        \noindent{\bf Appendix \thesubappendixc. {\kern1pt \bfit #1}}
	\par\vspace{5pt}}
\newcommand{\subsubappendix}[1] {\vspace{12pt}
        \refstepcounter{subsubappendixc}
        \noindent{\rm Appendix \thesubsubappendixc. {\kern1pt \tenit #1}}
	\par\vspace{5pt}}

\newcommand{\textlineskip}{\baselineskip=13pt}
\newcommand{\smalllineskip}{\baselineskip=10pt}

\def\eightcirc{
\begin{picture}(0,0)
\put(4.4,1.8){\circle{6.5}}
\end{picture}}
\def\eightcopyright{\eightcirc\kern2.7pt\hbox{\eightrm c}} 

\newcommand{\copyrightheading}[1]
	{\vspace*{-2.5cm}\smalllineskip{\flushleft
	{\footnotesize International Journal of Modern Physics A, #1}\\
	{\footnotesize $\eightcopyright$\, World Scientific Publishing
	 Company}\\
	 }}

\def\abstracts#1#2#3{{
	\centering{\begin{minipage}{4.5in}\baselineskip=10pt\footnotesize
	\parindent=0pt #1\par 
	\parindent=15pt #2\par
	\parindent=15pt #3
	\end{minipage}}\par}} 

\newcommand{\bibit}{\nineit}

\renewenvironment{thebibliography}[1]
	{\frenchspacing
	 \ninerm\baselineskip=11pt
	 \begin{list}{\arabic{enumi}.}
	{\usecounter{enumi}\setlength{\parsep}{0pt}
	 \setlength{\leftmargin 12.7pt}{\rightmargin 0pt} 
	 \setlength{\itemsep}{0pt} \settowidth
	{\labelwidth}{#1.}\sloppy}}{\end{list}}

\newcounter{itemlistc}
\newcounter{romanlistc}
\newcounter{alphlistc}
\newcounter{arabiclistc}

\newenvironment{romanlist}
	{\setcounter{romanlistc}{0}
	 \begin{list}{$($\roman{romanlistc}$)$}
	{\usecounter{romanlistc}
	 \setlength{\parsep}{0pt}
	 \setlength{\itemsep}{0pt}}}{\end{list}}

\newcommand{\fcaption}[1]{
        \refstepcounter{figure}
        \setbox\@tempboxa = \hbox{\footnotesize Fig.~\thefigure. #1}
        \ifdim \wd\@tempboxa > 5in
           {\begin{center}
        \parbox{5in}{\footnotesize\smalllineskip Fig.~\thefigure. #1}
            \end{center}}
        \else
             {\begin{center}
             {\footnotesize Fig.~\thefigure. #1}
              \end{center}}
        \fi}

\newcommand{\tcaption}[1]{
        \refstepcounter{table}
        \setbox\@tempboxa = \hbox{\footnotesize Table~\thetable. #1}
        \ifdim \wd\@tempboxa > 5in
           {\begin{center}
        \parbox{5in}{\footnotesize\smalllineskip Table~\thetable. #1}
            \end{center}}
        \else
             {\begin{center}
             {\footnotesize Table~\thetable. #1}
              \end{center}}
        \fi}

\def\@citex[#1]#2{\if@filesw\immediate\write\@auxout
	{\string\citation{#2}}\fi
\def\@citea{}\@cite{\@for\@citeb:=#2\do
	{\@citea\def\@citea{,}\@ifundefined
	{b@\@citeb}{{\bf ?}\@warning
	{Citation `\@citeb' on page \thepage \space undefined}}
	{\csname b@\@citeb\endcsname}}}{#1}}

\newif\if@cghi
\def\cite{\@cghitrue\@ifnextchar [{\@tempswatrue
	\@citex}{\@tempswafalse\@citex[]}}
\def\citelow{\@cghifalse\@ifnextchar [{\@tempswatrue
	\@citex}{\@tempswafalse\@citex[]}}
\def\@cite#1#2{{$\null^{#1}$\if@tempswa\typeout
	{IJCGA warning: optional citation argument 
	ignored: `#2'} \fi}}

\def\pmb#1{\setbox0=\hbox{#1}
	\kern-.025em\copy0\kern-\wd0
	\kern.05em\copy0\kern-\wd0
	\kern-.025em\raise.0433em\box0}

\def\fnt#1#2{\footnotetext{\kern-.3em
	{$^{\mbox{\scriptsize #1}}$}{#2}}}

\def\fpage#1{\begingroup
\voffset=.3in
\thispagestyle{empty}\begin{table}[b]\centerline{\footnotesize #1}
	\end{table}\endgroup}

\def\runninghead#1#2{\pagestyle{myheadings}
\markboth{{\protect\footnotesize\it{\quad #1}}\hfill}
{\hfill{\protect\footnotesize\it{#2\quad}}}}
\font\tenrm=cmr10
\font\tenit=cmti10 
\font\tenbf=cmbx10
\font\bfit=cmbxti10 at 10pt
\font\ninerm=cmr9
\font\nineit=cmti9

\font\eightrm=cmr8

\def\qed{\hbox{${\vcenter{\vbox{			
   \hrule height 0.4pt\hbox{\vrule width 0.4pt height 6pt
   \kern5pt\vrule width 0.4pt}\hrule height 0.4pt}}}$}}

\begin{document}

\runninghead{Galaxy Dynamics and the Second Peak} {Cold Dark Matter?}

\normalsize\textlineskip
\thispagestyle{empty}
\setcounter{page}{1}

\copyrightheading{}			

\vspace*{0.88truein}

\fpage{1}
\centerline{\bf DYNAMICS AND THE SECOND PEAK: COLD DARK MATTER?}
\vspace*{0.035truein}
\centerline{\footnotesize STACY MCGAUGH}
\vspace*{0.015truein}
\centerline{\footnotesize\it Department of
Astronomy, University of Maryland}
\baselineskip=10pt
\centerline{\footnotesize\it College Park, Maryland 20742-2421, USA}
\vspace*{10pt}

\vspace*{0.21truein}
\abstracts{The amplitude of the second peak in the angular power
spectrum of the cosmic microwave background radiation
is constrained to be small by recent experiments like
Boomerang.  This is surprising in the context of the
$\Lambda$CDM model, which predicted a large second 
peak.  However, this result is
expected if CDM does not exist.  The observed shape
of the power spectrum was accurately predicted
(before the fact) by a model motivated by the
surprising recent successes of the modified dynamics
(MOND) hypothesized by Milgrom.}{}{}


\vspace*{1pt}\textlineskip	
\section{Why CDM?}
\vspace*{-0.5pt}
\noindent
Fundamental to modern cosmology is the notion that the bulk of the mass of
the universe is composed of some non-baryonic form of matter generically
referred to as Cold Dark Matter$^1$ (CDM).  
The existence of CDM is inferred for several reasons:
\begin{romanlist}
\item $\Omega_m \gg \Omega_b$.
\item The quick growth of structure from smooth at
	$z \sim 1000$ to lumpy at $z = 0$.
\item The desire for $\Omega_m =1$, the only natural scale in FRW models.
\end{romanlist}

The critical assumption which underlies all of these is
that standard gravitational theory is completely adequate
on the scales relevant to galaxies and cosmology.
On the one hand, this seems like the safest possible assumption.
On the other hand, it manifestly fails in galaxies and clusters of galaxies.
These systems exhibit clear mass discrepancies; it is our faith in the
inviolability of the inverse square law that leads us to infer the existence
of dark matter.

A question which might be posed to distinguish these two reasonable but
opposite attitudes is:  {\it Can a single force law explain the observed
mass discrepancies, based only on the luminous matter?}  If dark matter is
the answer, then there should be no direct way to predict motions based only
on the luminous matter.  If a modified form of the force law comes into effect
on some galactic scale, then it must apply everywhere.

Surprisingly, there is one such idea which
has had considerable success both in explaining and predicting astronomical
observations.  This is the modified Newtonian dynamics (MOND) suggested by
Milgrom.$^2$  MOND supposes that for accelerations $a \ll a_0 \approx
1.2 \times 10^{-10}\;{\rm m}\,{\rm s}^{-2}$, the effective acceleration
becomes $a \rightarrow \sqrt{g_N a_0}$, where $g_N$ is the usual Newtonian
acceleration which applies when $a \gg a_0$.  
MOND has had considerable success in predicting the dynamics of a
remarkably wide variety of objects.$^3$  These include spiral galaxies,
low surface brightness galaxies,
dwarf spheroidals, giant ellipticals, groups and clusters of galaxies,
and large scale filaments.
Though perhaps not widely known, the dynamical evidence which supports
MOND is quite strong.  At the least, it constitutes an observed
phenomenology for which there is no plausible explanation in the 
context$^4$ of CDM.

\begin{figure}[h]
\vspace*{13pt}
\centerline{\vbox{\hrule width 5cm height0.001pt}}
\vspace*{13pt}
\centerline{{\epsfig{file=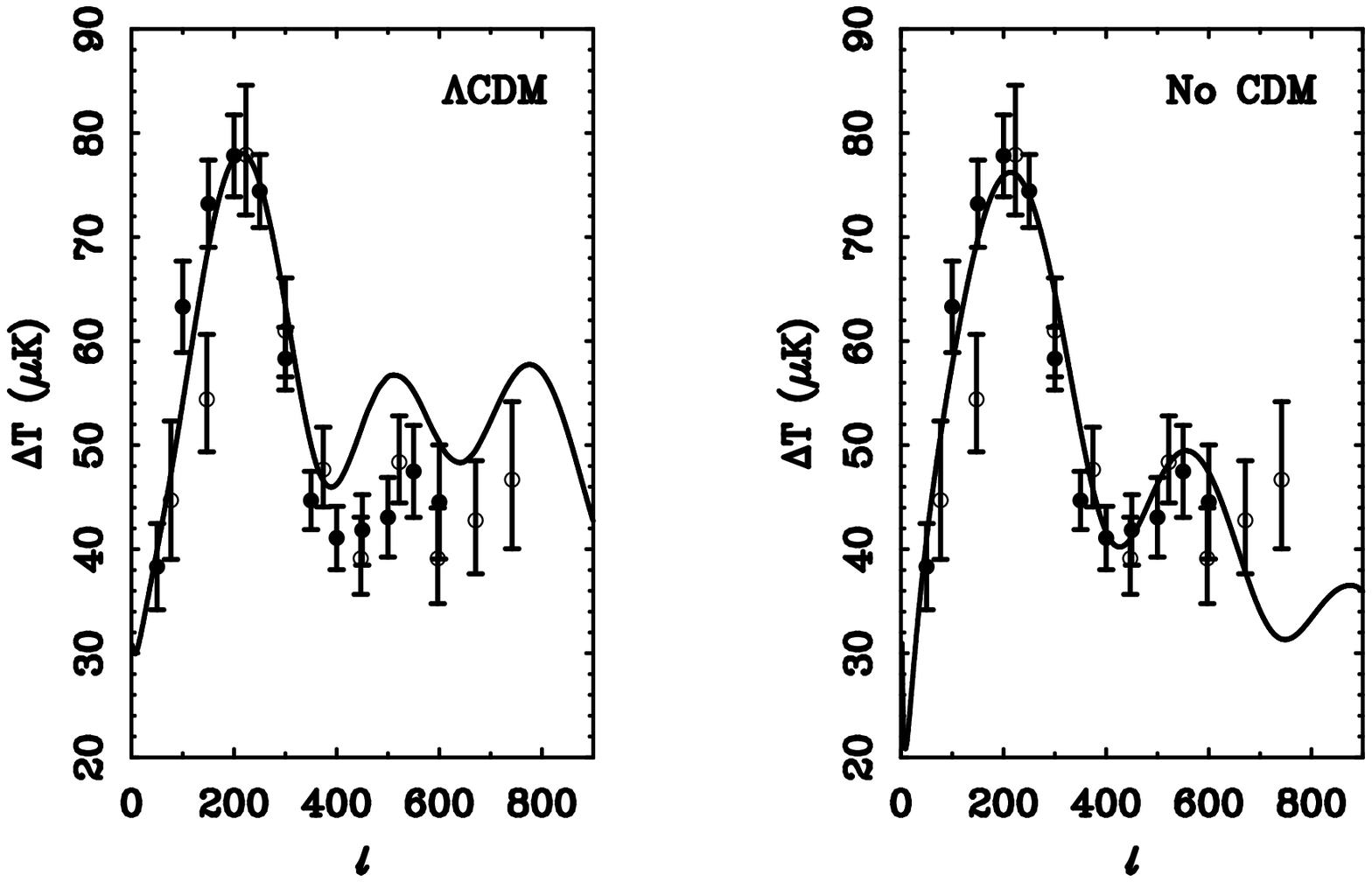,width=5truein}}}
\centerline{\vbox{\hrule width 5cm height0.001pt}}
\vspace*{13pt}
\fcaption{The angular power spectrum of fluctuations in the microwave
background, as observed and predicted.  The data are from the Boomerang$^{5}$
(solid points) and Maxima-1$^{6}$ (open points) experiments.  The calibration
offset between the two have been reconciled by scaling the Boomerang data
up by 13.5\%, the amount which minimizes the difference in the shape
of the spectrum.  In the left panel is the $\Lambda$CDM model as it existed
prior to Boomerang.  It is inconsistent with these data at greater than the
$99\%$ confidence level.
In the right panel, the line shows a purely baryonic model
which succeeded$^{7}$ in predicting$^{8}$ the shape of the power
spectrum {\it a priori}.  The only free parameters in this fit are the
amplitude $\Delta$T (unavoidable in any model) and the geometry place holder
$\Lambda$.  The baryon density which gives the best fit
($\chi_{\nu}^2 = 0.85$) to the data is
exactly that indicated by big bang nucleosynthesis
and the observed abundances of the light elements.$^{9}$
Parameters of the models are given in Table 1.}
\end{figure}

MOND does a plausible job of explaining items (i) -- (iii) which
motivated CDM.  The dynamical mass is overestimated when purely Newtonian
dynamics is employed in the MOND regime, so rather than $\Omega_m > \Omega_b$
one infers $\Omega_m \approx \Omega_b$.$^{4,10}$
The early universe is dense,
so accelerations are high and MOND effects do not appear until
after recombination.  When they do,
structure grows more rapidly than with conventional gravity$^{10}$
so the problem in going from a smooth
microwave background to a rich amount of large scale structure is also
alleviated.  Since everything is normal in the high acceleration regime,
all the usual early universe results are retained.  In addition, $\Omega_m =1$
ceases to be special.  In the absence of repulsive forces ($\Lambda$),
the universe will eventually recollapse$^{11}$ for any $\Omega_m$.

\section{Anisotropies in the Microwave Background}
\noindent
A test which goes beyond the dynamical succeses of MOND is
offered by the microwave background.
If a modified force law like MOND is the cause of the observed
mass discrepancies, CDM becomes unnecessary. 
Recent and upcoming experiments to measure the angular power spectrum of
anisotropies in the microwave background have the ability to distinguish
cosmologies with and without CDM.

In order to explore the expected difference
before the data came in, I considered$^{8}$ conventional $\Lambda$CDM models
and models with $\Omega_{CDM} = 0$.  The biggest difference between these is
in the amplitude of the peaks in the power spectrum subsequent to the first.
Without CDM, baryonic drag suppresses the power at progressively
smaller angular scales.  This is reflected in the low limit currently
placed on the amplitude of the second peak (Fig.~1).  This effect should
become more pronounced as the data improve and push to higher $\ell$.

\begin{table}[h]
\tcaption{Parameters of the Models in Fig.~1.}
\centerline{\footnotesize\smalllineskip
\begin{tabular}{l c c c c c}\\
\hline
{} &$\Omega_{tot}$ &$\Omega_b$ &$\Omega_{CDM}$ &$\Omega_{\Lambda}$ &$H_0$\\
\hline
$\Lambda$CDM & 1.0\phantom0 & 0.039 & 0.317 & 0.644 & 70 \\
No CDM & 1.04 & 0.034 & 0.0\phantom0\phantom0 & 1.006 & 75 \\
\hline\\
\end{tabular}}
\end{table}

The predictions of the purely baryonic case$^{8}$ have been realized$^{7}$
in the data subsequently reported by the Boomerang$^{5}$ and Maxima-1$^{6}$
experiments.  The spectrum has precisely the shape expected given known
values for all the cosmological parameters and $\Omega_{CDM} = 0$.
This, like the dynamical data before it,$^{3,4}$ suggests that CDM does
not exist.

\nonumsection{Acknowledgements}
\noindent
The work of SSM is supported in part by NSF grant AST 99-01663.

\nonumsection{References}


\begin{thebibliography}{000}
\bibitem{1}
T. A. Shutt {\bibit et al.} (2000), these proceedings.

\bibitem{2}
M. Milgrom, {\bibit Astrophys. J.} (1983), Vol. 270, p. 371.

\bibitem{3}
See http://www.astro.umd.edu/\~{}ssm/mond/litsub.html.

\bibitem{4}
S. S. McGaugh and W. J. G. de Blok, {\bibit Astrophys. J.} (1998), Vol. 499,
	p. 66.

\bibitem{5}
P. de Bernardis {\bibit et al.}, {\bibit Nature} (2000), Vol. 404, p. 955

\bibitem{6}
S. Hanany {\bibit et al.} (2000), astro-ph/0005123

\bibitem{7}
S. S. McGaugh, {\bibit Astrophys. J.} (2000), Vol. 541, p. L33

\bibitem{8}
S. S. McGaugh, {\bibit Astrophys. J.} (1999), Vol. 523, p. L99

\bibitem{9}
D. Tytler, J. M. O'Meara, N. Suzuki, and D. Lubin, {\bibit Physica Scripta}
	(2000), Vol. T 85, p. 12

\bibitem{10}
R. H. Sanders, {\bibit Mon. Not. R. Astr. Soc.} (1998), Vol. 296, p. 1009.

\bibitem{11}
J. E. Felten, {\bibit Astrophys. J.} (1984), Vol. 286, p. 3

\end{thebibliography}
\end{document}